\documentclass[twocolumn,a4,10pt]{article}
\usepackage{iasted}
\usepackage{times}
\usepackage{bbm}
\usepackage[dvips]{graphicx}
\usepackage{bm,amsfonts,amssymb,amsmath}
\usepackage[colorlinks=true,urlcolor=red]{hyperref}
\usepackage{graphicx}
\newcommand{\bth}{\begin{theorem}}
\newcommand{\enth}{\end{theorem}}

\newcommand{\bp}{\begin{proposition}}
\newcommand{\ep}{\end{proposition}}

\newtheorem{corrollary}{Corrollary}%[Ccor]
\newcommand{\bcor}{\begin{corrollary}}
\newcommand{\ecor}{\end{corrollary}}
\newcommand{\blem}{\begin{lemma}}
\newcommand{\elem}{\end{lemma}}
\newcommand{\bpb}{\begin{question}}
\newcommand{\epb}{\end{question}}

\newcommand{\bdf}{\begin{definition}}
\newcommand{\edf}{\end{definition}}
\newcommand{\bpr} {\begin{proof}}
\newcommand{\epr}{\qed \end{proof}}

\newcommand{\bd}{\begin{document}}
\newcommand{\ed}{\end{document}}
\newcommand{\beq}{\begin{equation}}
\newcommand{\eeq}{\end{equation}}
\newcommand{\bef}{\begin{figure}}
\newcommand{\enf}{\end{figure}}
\newcommand{\bea}{\begin{eqnarray}}
\newcommand{\eea}{\end{eqnarray}}
\newcommand{\baR}{\begin{array}}
\newcommand{\eaR}{\end{array}}
\newcommand{\bc}{\begin{center}}
\newcommand{\ec}{\end{center}}
\newcommand{\ben}{\begin{enumerate}}
\newcommand{\een}{\end{enumerate}}
\newcommand{\bit}{\begin{itemize}}
\newcommand{\eit}{\end{itemize}}
\newcommand{\su}{\section}
\newcommand{\ssu}{\subsection}
\newcommand{\sssu}{\subsubsection}
\newcommand{\nid}{\noindent}
\newcommand{\nnb}{\nonumber}

\newcommand\be{{\bf e}}
\newcommand\bv{{\bf v}}
\newcommand\h{{\bf h}}
\newcommand\bom{\mbox{{\boldmath $\omega$}}}
\newcommand\tom{\omega}
\newcommand\cM{{\cal M}}
\newcommand\cE{{\cal E}}
\newcommand\hM{{\hat{\cal M}}}
\newcommand\Mom{\cM_{\omega}}
\newcommand\on{\seq{\omega}{0}{n}}
\newcommand\onm{\seq{\omega}{0}{n-1}}
\newcommand\onmd{\left[\omega\right]_{n-2}}
\newcommand\ot{\left[\omega\right]_t}
\newcommand\otm{\left[\omega\right]_{t-1}}
\newcommand\Mon{\cM_{\on}}
\newcommand\Monm{\cM_{\onm}}
\newcommand\hMon{\hat{\cM}_{\on}}
\newcommand\hMot{\hat{\cM}_{\ot}}
\newcommand\oMon{\stackrel{\circ}{\cM}_{\on}}
\newcommand\ohMon{\stackrel{\circ}{\hM}_{\on}}
\newcommand\Anion{A_{i,n,\on}}
\newcommand\bAnion{\bar{A}_{i,n,\on}}
\newcommand\Fon{\F_{\on}}
\newcommand\Fot{\F_{\ot}}
\newcommand\hF{\hat{\F}}
\newcommand\hFon{\hF_{\on}}
\newcommand\hFot{\hF_{\ot}}
\newcommand\V{{\bf V}}
\newcommand\sV{\bf{\tiny{V}}}
\newcommand\f{{\bf{f}}}
\newcommand\hV{{\hat{\V}}}
\newcommand\hS{{\hat{\cS}}}
\newcommand\hml{{\hat{\mu}_0}}
\newcommand\piu{\pi^{(u)}}
\newcommand\pis{\pi^{(s)}}
\newcommand\tin{\tau_i(\omega,n)}

\newcommand\cEta{\cC(\bEta)}

\newcommand\bEtap{\mbox{{\boldmath $\eta'$}}}
\newcommand\cEtap{\cC(\bEtap)}
\newcommand\betat{\tilde{\bEta}}
\newcommand\betatp{\tilde{\bEtap}}
\newcommand\betatpl{\tilde{\bEta}^+}
\newcommand\bbeta{\mbox{{\boldmath $\beta$}}}
\newcommand\mub{\mu_{\bbeta}}
\newcommand\mus{\mu^\ast}
\newcommand\Ps{P^\ast}
\newcommand\Fb{\cF_{\bbeta}}
\newcommand\bxi{\bm{\xi}}
\newcommand\balpha{{\bm{\alpha}}}
\newcommand\blambda{{\bm{\lambda}}}
\newcommand\bphi{{\bm{\phi}}}
\newcommand\B{{\bf B}}
\newcommand\Vp{{\bf V'}}
\newcommand\W{{\bf W}}
\newcommand\Z{{\bf Z}}
\newcommand\F{{\bf F}}
\newcommand\Fe{\F_{\bEta}}
\newcommand\Fei{F_{\bEta,i}}
\newcommand\G{{\bf G}}
\newcommand\I{{\bf I}}
\newcommand\bz{{\bf 0}}
\newcommand\bzs{\left\{\bz\right\}}
\renewcommand\S{{\bf S}}
\newcommand\Td{T\left(\dAS\right)}
\newcommand\X{{\bf X}}
\newcommand\x{{\bf x}}
\newcommand\y{{\bf y}}
\newcommand\xt{\X_t(\V)}
\newcommand\cktv{\chi_{k,t}(\V)}
\newcommand\cA{{\cal A}}
\newcommand\cB{{\cal B}}
\newcommand\cH{{\cal H}}
\newcommand\cL{{\cal L}}
\newcommand\cN{{\cal N}}
\newcommand\cO{{\cal O}}
\newcommand\bme{\cB_\cM(\epsilon)}
\newcommand\Bev{\cB(\V,\epsilon)}
\newcommand\coMe{\stackrel{\circ}{\cMe}}
\newcommand\beV{\cB(\V,\epsilon)}
\newcommand\bmd{\cB_\cM(\delta)}
\newcommand\bmdp{\cB_\cM(\delta')}
\newcommand\cC{{\cal C}}
\newcommand\cD{{\cal D}}
\newcommand\cJ{{\cal J}}
\newcommand\cR{{\cal R}}
\newcommand\De{\cD(\bEta)}
\newcommand\bDe{\bar{\cD}(\bEta)}
\newcommand\Dep{\cD(\bEtap)}
\newcommand\bDep{\bar{\cD}(\bEtap)}
\newcommand\cG{{\cal G}}
\newcommand\GW{\cG_{(\cW,\Ie)}}
\newcommand\IW{\cI_{(\cW,\Ie)}}
\newcommand\cF{{\cal F}}
\newcommand\bcF{\bar{\cF}}
\newcommand\oGF{{\omega(\GF)}}
\newcommand\oGFTe{{\omega(\GFTe)}}
\newcommand\oF{{\omega_{\cF}(\GF)}}
\newcommand\oFTe{{\omega_{\cF}(\GFTe)}}
\newcommand\Bie{\cB_{\infty,\epsilon}}
\newcommand\BTe{\cB_{T,\epsilon}}
\newcommand\BTep{\cB_{{T+1},\epsilon}}
\newcommand\BTeu{\cB_{T,\epsilon}^{(1)}}
\newcommand\BTed{\cB_{T,\epsilon}^{(2)}}
\newcommand\DFFTV{\left. D\FTF\right|_{\V}}
\newcommand\FF{{\F_{\cF}}}
\newcommand\FTF{{\FT_{\cF}}}
\newcommand\FFb{{\F_{\bcF}}}
\newcommand\VF{{\V_{\cF}}}
\newcommand\VFb{{\V_{\bcF}}}
\newcommand\ft{\cF_t(\V)}
\newcommand\ftmu{\cF_{t-1}(\V)}
\newcommand\ftmk{\cF_{t-k}(\V)}
\newcommand\ftmkpu{\cF_{t-k+1}(\V)}
\newcommand\bft{\bar{\cF}_t(\V)}
\newcommand\bftmu{\bar{\cF}_{t-1}(\V)}
\newcommand\bftmk{\bar{\cF}_{t-k}(\V)}
\newcommand\bftmkpu{\bar{\cF}_{t-k+1}(\V)}
\newcommand\cMz{{\cal M_{\bf{0}}}}
\newcommand\cMu{{\cal M_{\bf{1}}}}
\newcommand\cMe{{\cM_{\bEta}}}
\newcommand\cMeu{{\cM_{{\bEta}_1}}}
\newcommand\cMet{{\cM_{{\bEta}_t}}}
\newcommand\cMeT{{\cM_{{\bEta}_T}}}
\newcommand\cMep{{\cM_{\bEtap}}}
\newcommand\cU{{\cal U}}
\newcommand\cS{{\cal S}}
\newcommand\cI{{\cal I}}
\newcommand\cP{{\cal P}}
\newcommand\cQ{{\cal Q}}
\newcommand\cV{{\cal V}}
\newcommand\cW{{\cal W}}
\newcommand\Vpi{V^+_i}
\newcommand\vp{\cV^+}

\newcommand\Is{\I^s}
\newcommand\IsP{\I^{s\ast}}
\newcommand\Ise{\I^s(\bEta)}
\newcommand\IE{\I(\bEta)}
\newcommand\Isv{\I^s(\V)}
\newcommand\Isi{I^s_i}
\newcommand\Isj{I^s_j}
\newcommand\Iie{I_i(\bEta)}
\newcommand\Iiep{I_i(\bEta')}
\newcommand\Ije{I_j(\bEta)}
\newcommand\Ijep{I_j(\bEta')}
\newcommand\Isie{\Isi(\bEta)}
\newcommand\Isje{\Isj(\bEta)}
\newcommand\Isjep{\Isj(\bEtap)}
\newcommand\Isiv{\Isi(\V)}
\newcommand\Isvt{I^s_i(\V(t))}
\newcommand\Ie{{\bf I^{ext}}}
\newcommand\Iei{I_i}
\newcommand\Iej{I^{ext}_j}
\newcommand\tk{\tau^{(k)}}
\newcommand\tkp{\tau^{(k+1)}}
\newcommand\toi{\tau^{(1)}_i(\V)}
\newcommand\toik{\tau^{(k)}_i(\V)}
\newcommand\toike{\tau^{(k)}_i(\betat)}
\newcommand\toikpe{\tau^{(k+1)}_i(\V)}
\newcommand\toikep{\tau^{(k+1)}_i(\betat)}
\newcommand\dik{\delta_i^{(k)}(\V)}
\newcommand\PrW{\cP_\cW}
\newcommand\PrI{\cP_\Ie}
\newcommand\PrWi{\cP_{\cW,\Ie}}
\newcommand\PW{\cP_{(\cW,\Ie)}}
\newcommand\RW{\cR_{(\cW,\Ie)}}
\newcommand\cPT{\cP^{(T)}}
\newcommand\cMT{\cM_{{\bEta_0} \dots {\bEta_T}}}
\newcommand\FT{\F^T}
\newcommand\Ft{\F^t}
\newcommand\Ftp{\F^{t+1}}
\newcommand\Fttk{\F^{t_k}}
\newcommand\Fmu{\F^{-1}}
\newcommand\Fmd{\F^{-2}}
\newcommand\FmT{\F^{-T}}
\newcommand\dAS{d(\oM,\cS)}
\newcommand\dSV{d(\tVp,\cS)}
\newcommand\trV{\left\{\V(t)\right\}_{t \geq 0}}
\newcommand\tV{\tilde{\V}}
\newcommand\tVp{\tilde{\V}^+}
\newcommand\tpi{t^{pre}_i}
\newcommand\tpj{t^{post}_j}
\newcommand\mW{m_{\cW}}
\newcommand\sW{\sigma_{\cW}}
\newcommand\mI{m_{\Ie}}
\newcommand\sI{\sigma_{\Ie}}
\newcommand\sw{\sigma_{(\cW,\Ie)}}
\newcommand\sB{\sigma_B^2}
\newcommand\sd{\sigma^2}
\newcommand\Cst{[\betat]_{s,t}}
\newcommand\defCst{\left\{\betat | \bEta_s=\as, \dots, \bEta_t = \at    \right\}}
\newcommand\as{\mbox{{\boldmath{$\alpha$}}}_s}
\newcommand\at{\mbox{{\boldmath{$\alpha$}}}_t}
\newcommand\moytpsi{\bar{\psi}_\cW}
\newcommand\sn{\left\{1 \dots N \right\}}
\newcommand\GF{\Gamma_\cF}
\newcommand\GFTe{\Gamma_{\cF,T,\epsilon}}
\newcommand\GFTep{\Gamma_{\cF,T,\epsilon'}}
\newcommand\GFTpe{\Gamma_{\cF,{T+1},\epsilon}}
\newcommand\PF{\Pi_{\cF}}
\newcommand\PbF{\Pi_{\bcF}}
\newcommand\oM{\Omega}

\newcommand\wsg{\cW^s(\V)}
\newcommand\wsl{\cW^s_{loc}(\V)}
\newcommand\wslf{\cW^s_{loc}(\F(\V))}
\newcommand\wse{\cW^s_{\epsilon}(\V)}
\newcommand\wset{\cW^s_{\epsilon}(\V(t))}
\newcommand\tjk{\tau_j^{(k)}}
\newcommand\tik{\tau_i^{(k)}}
\newcommand\tiu{\tau_i^{(1)}}
\newcommand\tikp{\tau_i^{(k+1)}}
\newcommand\Vr{V_{reset}}
\newcommand\Vm{V_{min}}
\newcommand\VM{V_{max}}
\newcommand\Vs{V^{+}}
\newcommand\Vsi{V_i^{+}}
\newcommand\SLp{\Sigma_\Lambda^+}
\newcommand\SL{\Sigma_\Lambda}
\newcommand\SW{\Sigma_{(\cW,\Ie)}}
\newcommand\SWp{\Sigma_{(\cW,\Ie)}^+}
\newcommand{\D}{\displaystyle}
\newcommand{\deq}{\stackrel {\rm def}{=}}
\newcommand{\peq}{\stackrel {\rm \mu p.s.}{=}}

\newcommand\cyl[3]{\left[{#1}\right]_{#2}^{#3}}
\newcommand\seq[3]{{#1}_{#2}^{#3}}
\newcommand\prodset[3]{{#1}_{#2}^{#3}}
\newcommand\tinm{\tau_i(\seq{\omega}{-\infty}{n-1})}
\newcommand\uom{\underline\omega}
\newcommand\uX{\underline X}
\newcommand\ucF{\underline \cF}
\newcommand\uomp{\underline\omega'}
\newcommand\tiz{\tau_i(\uom)}
\newcommand{\setZ}{\mathbbm{Z}}
\newcommand{\setN}{\mathbbm{N}}
\newcommand{\setR}{\mathbbm{R}}
\newcommand{\Exp}[1]{\mathbb{E}\left [ #1 \right]}
\newcommand{\ExpB}[1]{\mathbb{E}_B\left [ #1 \right]}
\newcommand{\CovB}[1]{Cov\left [ #1 \right]}
\newcommand\pTo{{\pi_{\omega}^{(T)}}}
\newcommand\tjn{t_j^{(n)}}
\newcommand\tot{\seq{\omega}{-\infty}{t}}
\newcommand\Jito{J_i(t,\tot)}

\begin{document}

\date{}

\title{Spike trains statistics in Integrate and Fire Models: exact results.}

\author{Bruno Cessac \url{http://www-sop.inria.fr/neuromathcomp/}, 
Hassan Nasser, 
Juan-Carlos Vasquez\thanks{INRIA NeuroMathComp}, 
} 

\maketitle

\thispagestyle{empty}

\noindent
{\bf\normalsize ABSTRACT}
We briefly review and highlight the consequences of
 rigorous and exact results obtained in \cite{cessac:10},
 characterizing the statistics of 
spike trains in a network of leaky Integrate-and-Fire neurons,
where time is discrete and where neurons are subject to noise,
without restriction on the synaptic weights connectivity. 
The main result is  that spike trains statistics are 
characterized by a Gibbs distribution, whose potential is explicitly computable. This establishes, on one hand, 
a rigorous ground for the current investigations attempting to characterize
real spike trains data with Gibbs distributions, such as the Ising-like 
distribution \cite{schneidman-berry-etal:06}, using the maximal entropy principle.
However, it transpires from the present analysis that the Ising model might be a rather weak
approximation. Indeed,  the Gibbs potential (the formal ``Hamiltonian'') 
 is the log of the so-called ``conditional intensity''
(the probability that a neuron fires given the past of the whole network \cite{johnson-swami:83,brillinger:88,chornoboy-schramm-etal:88,kass-ventura:01,truccolo-eden-etal:05,okatan-wilson-etal:05,truccolo-donoghue:07,pouzat-chaffiol:09}).
But, in the present example, this probability has an infinite memory, and the corresponding process is non-Markovian
(resp. the Gibbs potential has infinite range).
Moreover, causality implies that the conditional intensity does not depend on the
state of the neurons at the \textit{same time}, ruling out the Ising model
as a candidate for an exact characterization of spike trains statistics.
However, Markovian approximations can be proposed whose degree of approximation
can be rigorously controlled. In this setting, Ising model appears as the ``next step''
after the Bernoulli model (independent neurons) since it introduces spatial pairwise correlations,
but not time correlations. The range of validity of this approximation is discussed together
with possible approaches allowing to introduce time correlations, with algorithmic
extensions.

\vspace{2ex}
   

\section{Introduction.}

At first glance, the neuronal activity is manifested by the emission of 
action potentials (``spikes'') constituting spike trains. 
Those spike trains are usually not exactly reproducible
when repeating the same experiment,
even with a very good control ensuring that the experimental
conditions have not changed. Therefore,
 researchers are seeking statistical
regularities in order to provide
an accurate model for spike train statistics, i.e. a probability
distribution fitting the experimental data and/or 
matching what neuroscientists believe models should predict.
However, obtaining  statistical models from experimental data remains a difficult task.

It appears simpler to characterize spike trains statistics in 
neural networks \textit{models} where one controls exactly
the neural network parameters, the number of involved
neurons, the number of samples, and the duration of the 
experiment (with a possible mathematical extrapolation
to infinite time). Especially, producing analytical
(and when possible, rigorous) results on those statistics
provide clues toward resolving  experimental questions
and new algorithms for data treatments.
Here, we propose a complete and rigorous characterization of
spike train statistics for the discrete-time leaky Integrate-and-Fire  model with  noise and time-independent stimuli.
This framework affords extrapolation to more realistic neural networks models such as generalized
Integrate-and-Fire \cite{rudolph-destexhe:06,cessac-vieville:08}.
Our results hold for finite-sized
networks, and all type of
synaptic graph structure are allowed. Also, we are not constrained
by an ad hoc choice of the initial conditions distribution of membrane potentials;
 instead we propose a procedure where this distribution is selected by dynamics
and  is uniquely determined.
More precisely, we show
that spike train statistics are characterized by a (unique) invariant probability distribution
 (equilibrium state) whatever
the model-parameters values, which satisfies a variational
principle (the maximal entropy principle) and is a Gibbs distribution whose potential is explicitly computed. 
This has several deep consequences discussed in this paper.

\section{Model definition.} \label{Smodeldef}
 We consider a discrete-time (continuous state) Integrate and Fire model,
introduced in  \cite{soula-beslon-etal:06,cessac:08}, whose dynamics is given by:

\beq \label{DNN}
V_i(t+1)= \gamma V_i \left(1 - Z[V_i(t)] \right)+ \sum_{j=1}^N W_{ij}Z[V_j(t)]+ \Iei + \sigma_B B_i(t),
\eeq
\nid where $i=1 \dots N$ is the neuron index, $V_i$ the membrane potential of neuron $i$
(this is a continuous variable), $\gamma \in [0,1[$ is the leak rate,
$Z(x)=1$ whenever $x \geq \theta$ and $Z(x)=0$ otherwise, where $\theta$ is the firing threshold,
$W_{ij}$ is the synaptic weight from neuron $j$ to neuron $i$, $\Iei$ is a constant external current,
$\sigma_B > 0$, and $B_i(t)$ is a noise, namely the $B_i(t)$'s are Gaussian i.i.d. random variable
with mean zero and variance $1$. The term $\left(1 - Z[V_i] \right)$ corresponds to the reset
of membrane potential whenever neuron $i$ fires while the term $\sum_{j=1}^N W_{ij}Z[V_j]$
is the post synaptic potential generated by the pre-synaptic spikes.

\section{Spike train statistics.}
To each membrane potential value $V_i(t)$ we associate a variable $\omega_i(t)=Z(V_i(t))$.
The ``spiking pattern'' of the neural network at time $t$ is the vector 
$\omega(t)=\left(\omega_i(t)\right)_{i=1}^N$: it tells us which neurons are firing at time $t$,
($\omega_i(t)=1$) and which neurons are not firing at time $t$ ($\omega_i(t)=0$). 
We denote by $\seq{\omega}{s}{t}$ the sequence
or \textit{spike block } $\omega(s) \dots \omega(t)$. 
A bi-infinite sequence $\tom=\left\{\omega(t)\right\}_{t=-\infty}^{+\infty}$ 
of spiking patterns is called a 
 ``raster plot''. It tells us which neurons are firing at each 
time $t \in \setZ$. In practical experimental raster plots are
obviously finite sequences of spiking pattern but the extension
to $\setZ$, especially the possibility of considering an arbitrary
distant past (negative times) is a key tool in the present
work. We are seeking invariant probability distributions on the
set of raster plots, generated by the dynamical system (\ref{DNN}).
In the next sections we give our main results whose proofs can be found
in \cite{cessac:10}.

\section{The transition probability.}
In this model it is possible to compute exactly and rigorously
the probability, called \textit{conditional intensity}
in \cite{johnson-swami:83,brillinger:88,chornoboy-schramm-etal:88,kass-ventura:01,truccolo-eden-etal:05,okatan-wilson-etal:05,truccolo-donoghue:07,pouzat-chaffiol:09}, of a spiking pattern at time $t+1$,
$\omega(t+1)$, given the past of the network $\seq{\omega}{-\infty}{t}$.
To our knowledge this is the first time that such an exact and rigorous computation is achieved at the level of \textit{network}. It is given
by:

\beq\label{Pt}
P\left(\omega(t+1)| \seq{\omega}{-\infty}{t},\right)=\prod_{i=1}^N P\left(\omega_i(t+1)| \seq{\omega}{-\infty}{t}\right),
\eeq

\nid with

\beq\label{Pti}
\baR{lll}
P\left(\omega_i(t+1)| \seq{\omega}{-\infty}{t}\right)=\\
\omega_{i}(t+1)\pi\left(\frac{\theta-C_i(\seq{\omega}{-\infty}{t})}{\sigma_i(\seq{\omega}{-\infty}{t})}\right)+\\
\left(1-\omega_{i}(t+1)\right)\left(1-\pi\left(\frac{\theta-C_i(\seq{\omega}{-\infty}{t})}{\sigma_i(\seq{\omega}{-\infty}{t})}\right)\right),
\eaR
\eeq

\nid where 
$$C_i(\seq{\omega}{-\infty}{t})=
\sum_{j=1}^N W_{ij}x_{ij}(\seq{\omega}{-\infty}{t})+\Iei 
\frac{1-\gamma^{t+1-\tau_i({\seq{\omega}{-\infty}{t}})}}{1-\gamma},$$
with
$$x_{ij}(\seq{\omega}{-\infty}{t})=\sum_{l=\tau_i(\seq{\omega}{-\infty}{t})}^{t} \gamma^{t-l}\omega_j(l),$$
is the deterministic contribution of the network to the membrane potential of neuron $i$ at time $t+1$, when the 
spike train up to time $t$ is $\seq{\omega}{-\infty}{t}$. 
The term $C_i(\seq{\omega}{-\infty}{t})$ integrates the pre-synaptic flux that arrived at neuron $i$ in the past, as well
as the external current $\Iei$. As a consequence, this term, 
depends on the past history up to a time $\tau_i(\seq{\omega}{-\infty}{t})$
which is the last time before $t$ when neuron $i$ has fired, in the sequence $\seq{\omega}{-\infty}{t}$
(the mathematical definition is given below). 

Likewise, $$\sigma^2_i(\seq{\omega}{-\infty}{t})=\sigma_B^2\frac{1-\gamma^{2(t+1-\tau_i(\seq{\omega}{-\infty}{t}))}}{1-\gamma^2}$$ is the variance of the noise-induced term applied to neuron $i$ at time $t+1$
and resulting from the integration of the noise from time $\tau_i(\seq{\omega}{-\infty}{t})$ to time $t$.
Finally, $\pi(x)=\frac{1}{\sqrt{2\pi}}\int_x^{+\infty} e^{-\frac{u^2}{2}}du.$

The mathematical definition of $\tau_i(\seq{\omega}{-\infty}{t})$ is:
$$\tau_i(\seq{\omega}{-\infty}{t})
\deq
\left\{
\baR{lll}
-\infty, \quad \mbox{if} \quad \omega_i(k)=0, \quad \forall k \leq t;\\
\max \left\{-\infty < k \leq t,\omega_i(k)=1\right\} \quad \mbox{otherwise}.
\eaR
\right.
$$
Basically, this notion integrates the fact that the state of  neuron $i$ depends
on  spikes emitted in the past by the network, \textit{up to the last time when this neuron has fired} 
(thus, the present analysis  relies heavily on the structure of IF models where reset of the membrane potential
to a fixed value erases the past evolution).
Depending on the sequence $\seq{\omega}{-\infty}{t}$, this time can go arbitrary far in the past.
Now, here, sequences such that $\tau_i(\seq{\omega}{-\infty}{t})<-c$ for some positive $c$,
have a \textit{positive} probability (decaying exponentially fast with $c$).
The consequence is that the transition probability (\ref{Pt}) depends on the history
of the network on an \textit{unbounded} past. As a consequence dynamics is non Markovian.

Note also that (evidently ?) the structure of the neural network dynamics imposes \textit{causality}.
This appears explicitly in (\ref{Pti}) where the probability that neuron $i$ fires at time $t+1$
depends on spikes emitted \textit{up to time $t$}. In other words, this probability does
not depend on the spikes emitted by the other neurons \textit{at the same time} ($t+1$).
We shall come back to this remark when discussing the relevance of Ising model.
 
Finally, remark that the factorization of the probability (\ref{Pt})  expresses
the \textit{conditional} independence of spikes at given time. This is simply due to
the fact that
the only source of stochasticity, when past spikes are fixed (by the conditioning),
is the noise, which is assumed, in this model, to be independent between neurons.
But, (\ref{Pt}) does not imply that spikes are independent.\\

Let us emphasize what we have obtained. We have an explicit form for the probability
that a neuron fires given its past. From this we can obtain the probability of
any spike block. But, here there is technical difficulty, since
the corresponding stochastic process is not Markovian. Such process corresponds to 
a more elaborated concept called a \textit{chain with
compete connections}. As a consequence the existence and uniqueness of an
invariant measure is not given by  classical theorems
on Markov chains but requires more elaborated tools, widely developed by the community of stochastic
processes and ergodic theory (see \cite{maillard:07} for a nice introduction and useful references).
The price to pay is a rather abstract mathematical formalism but the reward is the possibility
of considering  spike statistics arising in such models, including memory effects and causality,
with, we believe, possible extensions toward more realistic neural models.

\section{Gibbs distribution.}
In the present setting where $\Iei$ does not depend on $t$, 
$P$ is stationary. Namely, fix a spiking sequence
$\left\{a_n\right\}_{n \leq 0}$, then, $\forall t$,
$P\left(\omega(t)=a_0 \, | \, \omega(t-n)=a_{-n}, \, n \geq 1 \right)=
P\left(a_0 \, | \, \seq{a}{-\infty}{-1}  \right).$
Therefore, instead of considering a family of
transition probabilities depending on $t$, it suffices to
define the transition probability 
at one time $t \in \setZ$, for example  $t=0$.
The next results are based on techniques from ergodic theory and
chains with complete connections \cite{maillard:07}.\\

We recall first a few definitions. A \textit{Gibbs distribution} is a probability
distribution $\mu_\psi$, on the set of (infinite) spike sequences (raster plots)
$X$, 
such  that one can find some constants 
$P(\psi),c_1,c_2$ with $0 < c_1 \leq 1 \leq c_2$ such
that for all $n \geq 0$ and for all $\omega \in X$:
$$
c_1 \leq \frac{\mu_\psi\left(\seq{\omega}{0}{n}\right)}
{\exp \left[ -(n+1) P(\psi)+\sum_{k=0}^n\psi(T^k\omega) \right]}  \leq c_2.$$
where $T$ is the \textit{right} shift\footnote{The use of the right shift instead of the left shift is standard in the 
context of chains with complete connections \cite{maillard:07}.} over the set of infinite sequences 
$X$ i.e. $(T\omega)(t)=\omega(t-1)$.
$P(\psi)$, called the topological pressure, is formally analogous to the free energy
and is a central quantity since it allows the computation of averages with respect to $\mu_\psi$.
In the previous equation, the term $\sum_{k=0}^n\psi(T^k\omega)$ may be viewed
as a formal Hamiltonian over a finite chain of symbols $\seq{\omega}{0}{n}$.
However, as exemplified below (eq. (\ref{psi})) this term depends also on the
past (infinite chain $\seq{\omega}{-\infty}{-1}$). There is an analogy with statistical physics
systems with boundary conditions where the Gibbs distribution not only depends on the configuration
inside the lattice, but also on the boundary term \cite{ruelle:78}.  

Moreover, $\mu_\psi$ is an \textit{equilibrium state} if it 
 satisfies (``maximum entropy principle''):
\beq \label{VarPrinc}
P(\psi) \deq h(\mu_\psi)+\mu_\psi(\psi)=\sup_{\mu \in P_T(X) } h(\mu)+\mu(\psi),
\eeq
where $P_T(X)$ the set of  $T$-invariant finite measures on $X$,
$h(\mu_\psi)$
is the entropy of $\mu_\psi$, and $\mu(\psi) \deq \int \psi d\mu$ is the average value
of $\psi$ with respect to the measure $\mu$. Though Gibbs states and equilibrium states are non equivalent notions
in the general case,
they are equivalent in the present setting.

Now, our main result is:\\

\textit{The dynamical system (\ref{DNN}) has a unique 
invariant-measure, $\mu_\psi$, whatever the values of parameters
$W_{ij}, \, i,j=1 \dots N, I_i, \, i=1 \dots N, \gamma, \theta$.
This is a Gibbs distribution and an equilibrium state, for the potential:
\beq\label{psi}
\baR{lll}
\psi(\omega) = \log(P\left(\omega(0)| \seq{\omega}{-\infty}{-1}\right)) \equiv \psi(\seq{\omega}{-\infty}{0})\\
 =\sum_{i=1}^N 
\left[\omega_{i}(0) 
\log\left(\pi\left(\frac{\theta-C_i(\uom)}
{\sigma_i(\uom)}\right)\right)\right.+ \\
\left.\left(1-\omega_{i}(0)\right)
\log
\left(1-\pi\left(\frac{\theta-C_i(\uom)}{\sigma_i(\uom)}\right)\right)
\right],
\eaR
\eeq
where we note $\uom \deq \seq{\omega}{-\infty}{-1}$.}

\bigskip

Note therefore that the Gibbs potential is simply the log of the conditional intensity.
We believe that this last remark extends to more general models.

 Knowing that $\mu_\psi$ is Gibbs distribution with a known potential
allows one to control the probability distribution of finite sequences.
Note that here the potential $\psi$ is \textit{explicitly} known.
This has several other deep consequences. First,
this formalism opens up the possibility of characterizing $P(\psi)$ and $\mu_\psi$
by a spectral approach, being respectively the (unique) largest
eigenvalue and related left eigenfunction of the Ruelle-Perron-Frobenius
operator $\cL_{\psi} f(\omega)=\sum_{\omega' \, : \, T(\omega')=\omega} \psi(\omega')f(\omega')$,
 acting on  $C(X,\setR)$,  the set of continuous real functions on $X$.
This last property has deep consequences at the implementation level \cite{vasquez-cessac-etal:10}.

Also, it is possible to propose Markovian approximations of this distribution, whose
degree of accuracy can be controlled, as we now discuss.

\section{Markov approximations.}
The main difficulty in handling the transition probabilities (\ref{Pt}) and the related equilibrium state
is that they depend on an history dating back to 
 $\tau_i(\seq{\omega}{-\infty}{t})$, where $\tau_i(\seq{\omega}{-\infty}{t})$ is unbounded.
On the other hand, the influence of 
the activity of the network, say  at time $-l$, on the 
membrane potential $V_i$ at time $0$, appearing
in the term 
$x_{ij}(\seq{\omega}{-\infty}{0})=\sum_{l=\tau_i(\seq{\omega}{-\infty}{0})}^{0} \gamma^{-l}\omega_j(l),$
 decays exponentially fast as $l \to -\infty$.
 Thus, one
may argue that after a characteristic time depending
on $\frac{1}{|\log(\gamma)|}$
the past network  activity has little influence on the current
membrane potential value.

Assume that we want to approximate the statistics of
spikes, given by the dynamics (\ref{DNN}), by  fixing a finite time horizon
$R$ such that  the membrane potential at time $0$  depends on the past only up to some
finite time $-R$. 
In this way, we truncate the histories and we approximate the transition probabilities 
$P\left(\omega(0)\, | \, \seq{\omega}{-\infty}{-1}\right)$,
 with unbounded memory,
by transition probabilities $P\left(\omega(0)\, | \, \seq{\omega}{-R}{-1}\right)$,
 thus limiting memory to at most $R$ time steps in the past.
These approximated transition probabilities
 constitute therefore a Markov chain with a memory depth $R$. 
But how good is this approximation ?

The truncation of histories leads to a truncation of the Gibbs potential, denoted $\psi^{(R)}$.
The invariant measure of the Markov chain is a Gibbs distribution for the
potential $\psi^{(R)}$, denoted $\mu_{\psi^{(R)}}$. One can show that
the Kullback-Leibler divergence between the exact measure $\mu_\psi$
and the Markov measure $\mu_{\psi^{(R)}}$ obeys: 
\beq\label{dKL}
d\left(\mu_{\psi^{(R)}},\mu_\psi \right) < C \gamma^R,
\eeq
where $C$ can be computed explicitly as a function of synaptic weights and current.

Therefore, the Kullback-Leibler divergence  decays exponentially fast, with a decay rate $\gamma$,
as expected from our prior qualitative analysis. It is thus sufficient, for practical purposes, to approximate
$\psi$ with a potential of range:
\beq\label{RApprox}
R \sim \frac{1}{|\log \gamma|}.
\eeq
Consequently, the smaller the leak, the shorter is the range.

Note that the previous result is classical
in ergodic theory and expresses that finite range potentials are dense in the set of Gibbs potentials
\cite{ruelle:78}. What we bring is the computation of the decay rate and an estimation of the constant $C$
for the present model (see \cite{cessac:10} for details).

\section{Raster plots statistics}
As discussed in the introduction, the neuroscience community is confronted to the delicate
problem of characterizing statistical properties of
raster plots from finite time spike trains and/or from finite number
of experiments. This requires an a priori guess
for the probability of raster plots, what we call 
a \textit{statistical model}. These models
can be extrapolated from methods
 such as Jaynes' \cite{jaynes:57}:
``Maximising the statistical entropy under constraints''.
Actually, this is none other that the variational
formulation (\ref{VarPrinc}). Now, this method
only provides an approximation of the sought probability,
relying heavily on the choice of constraints usually fixed from phenomenological arguments,
and this approximation can be rather bad  \cite{csiszar:84}.
Moreover, note that Jaynes' method is typically used for a finite number of constraints
while (\ref{VarPrinc})
holds for a very general potential, actually corresponding, in
the present context, to infinitely many constraints.

On phenomenological grounds, spike $n$-uplets $(i_1,t_1), \dots, (i_n,t_n)$ (neuron $i_1$
fires at time $t_1$, and neuron $i_2$ fires at time $t_2$,
$\dots$) provide natural choices of constraints since they correspond
to experimentally measurable events whose probability of occurrence has a phenomenological
relevance. For example, the  probability of occurrence of  $(i_1,t_1)$,
$\mu_\psi(\omega_i1(t_1))$, 
is the instantaneous firing rate of neuron $i_1$ at time $t_1$.
 On a more formal ground a spike $n$-uplet corresponds to
 an \textit{order-$n$ monomial}; this is
a function $\phi$ which associate to a raster $\omega$
the product $\omega_{i_1}(t_1) \dots \omega_{i_n}(t_n)$,
where $1 \leq i_1 \leq i_2 \leq \dots \leq i_n \leq N$ and
$-\infty < t_1 \leq t_2 \leq \dots \leq t_n < 0$, and where there is
no repeated pair of indexes $(i,t)$. Hence, $\phi(\omega)$ is equal to $1$ if and only if neuron $i_1$
fires at time $t_1$, and neuron $i_2$ fires at time $t_2$,
$\dots$ in the raster $\omega$ (and is $0$ otherwise). A \textit{polynomial} is a linear combination of monomials.

Fixing constraint on spike-uplets (monomials) average leads to
specific forms of Gibbs distribution. As an example, constraining firing rates
only corresponds to a Bernoulli distribution where
neurons spike independently and where the potential reads $\sum_{i=1}^N \lambda_i \omega_i(0)$. Constraining
rates and probability that 2 neurons fire simultaneously leads to the
so-called Ising distribution where the potential reads $\sum_{i=1}^N \lambda_i \omega_i(0)+
\sum_{i,j=1}^N \lambda_{ij} \omega_i(0)\omega_j(0)$ \cite{schneidman-berry-etal:06}.
Here, the related probability measure does not factorize any more
but all information about spike train statistics is contained
in the first and second order spike-uplets at time $0$.

More generally, imposing  constraint on a (finite) set of monomials $\phi_1, \dots, \phi_L$, leads
to a parametric form of  Gibbs potential $\phi_{guess}(\omega) = \sum_{l=1}^L \lambda_l \phi_l(\omega)$,
where the $\lambda_l$ are Lagrange multipliers. Note that the $\lambda_l$'s are related by an additional constraint,
the normalisation of the probability. Again, this potential relies heavily on the choice of constraints.\\

On the opposite, in the present example, instead of an ad hoc guess,
 an exact polynomial expansion can be obtained from the explicit form of the
potential (\ref{psi}). It is given by expanding (\ref{psi}) in series via the series
expansion of $\log(\pi(\frac{\theta-C_i(\uom)}{\sigma_i(\uom)}))$
(note that $\frac{\theta-C_i(\uom)}{\sigma_i(\uom)}$ is bounded in absolute value provided the synaptic
weights are finite and the noise intensity $\sigma_B $ is positive. Thus, 
$0 < a < \pi(\frac{\theta-C_i(\uom)}{\sigma_i(\uom)}) < b<1$, for some $a,b$). This series involves
terms of the form $C_i(\uom)^n$ leading to monomials of the form $\omega_{j_1}(t_1) \dots
\omega_{j_n}(t_n)$ where $j_1, \dots, j_n$ are neurons pre-synaptic to $i$ (with $W_{i{j_k}} \neq 0, \dots k=1 \dots n$)
and $t_k <0, k=1 \dots n$. As a consequence, the potential (\ref{psi})
has the form $\sum_{i=1}^N \sum_{l=1}^\infty \lambda_{i,l} \phi_{i,l}(\omega)$
where the $\phi_{i,l}$'s are monomials of the form $\omega_i(0)\omega_{j_1}(t_1) \dots
\omega_{j_n}(t_n), \quad t_1, \dots t_n < 0 $ and 
 the $\lambda_{i,l}$'s contains products of the form 
$\gamma^{t_1} \dots \gamma^{t_n} W_{i{j_1}} \dots W_{i{j_n}}$. 

Note that
terms of the form $\omega_{i_1}(0)\omega_{i_2}(0) \dots$ do not appear in this expansion.
Especially, it \textit{does not contain the Ising term}.  

As discussed in the previous section, truncating this series to a finite polynomial
involving monomials with a time depth $R$ (i.e. of the form $\omega_{j_1}(t_1) \dots
\omega_{j_n}(t_n)$ with $- 1 \leq t_k \leq -R$) amounts to considering Markovian
approximations of $\psi$. How the corresponding 
Gibbs distribution approximates the exact one is controlled by eq. (\ref{dKL})
where $C$ depends on the synaptic weights and currents.

This expansion contains an exponentially increasing number of terms as $R$, or $N$,
the number of neurons,
growths. However, it does not contain all possible monomials. Indeed, beyond,
the remark above about the vanishing of non causal term 
the form of the potential  can be considerably reduced from elementary considerations
such as stationarity. Moreover, some $\lambda^{(l)}_{i_1,t_1,\dots,i_l,t_l}$'s can be
quite close to zero and eliminating them does not increase significantly the Kullback-Leibler
divergence. This can be used to perform systematic (algorithmic) reduction of the potential in a more general context than the present model (see next section and \cite{vasquez-cessac-etal:10} for more details). 

To summarize, one can obtain, from the previous analysis, a
canonical form for a range-$R$ approximation of $\psi$,
of the form:
\beq\label{gener_psi_R}
\psi_\blambda^{(R)}=\sum_{l=0}^K \lambda_l \phi_l,
\eeq
where all terms contribute significantly to the probability distribution
i.e. removing them leads to a significant increase of the KL divergence between
the exact Gibbs distribution and its approximation. Thus, in the present case, the relevant constraints
(and the value of the related $\lambda_l$'s) can be estimated from the analytic form of the potential.

\su{Parametric estimation of spike trains statistics.}
 This analysis opens up the possibility of developing efficients algorithms
to estimate at best the statistic of spike trains from experimental data, using several guess potential and selecting the one which minimizes the KL divergence between the Gibbs measure $\mu_{\psi_\blambda^{(R)}}$
and the empirical measure attached to some experimental raster $\omega^{(exp)}$ \cite{vasquez-cessac-etal:10}.
The idea is to start from a parametric form of potential (\ref{gener_psi_R}), of range $R$,
and to compute the empirical average of all monomials $\phi_l$ from the experimental raster $\omega^{(exp)}$.
Then, one adjust the parameters $\lambda_l$ by minimizing the KL divergence.
This computation can be easily done using spectral properties of the Perron-Frobenius
operator. This algorithm  described in \cite{vasquez-cessac-etal:10}, will be presented in this conference,
in another communication,
and is freely available as a C++ source at \url{http://enas.gforge.inria.fr}.

\section{Discussion}

In this paper we have addressed the question of characterizing the spike trains statistics of a network of LIF neurons with noise, in the stationary case,
 with two aims. Firstly, to obtain analytic and rigorous results allowing the characterization of the process of spike generations. Secondly,  to make a connection from this mathematical analysis toward the empirical methods used in neuroscience community for the analysis of spike trains. Especially, we have shown that the  ``Ising model'' provides, in this example,
 a bad approximation of the exact Gibbs potential. This is due, on one hand, to the fact that Ising potential does not take into account causality, and on the other hand, to the fact that the exact potential includes quite a bit more ``constraints'' than the mere average value of pairwise spike coincidence. Although the first objection can somewhat be relaxed when considering data binning (that we do not consider here), the second one seems unavoidable.  

Also, we have shown that the Jaynes method, based on an a priori choice of a ``guess'' potential, with finite range, amounts to approximate the exact probability distribution by the Gibbs distribution of a Markov chain. The degree of approximation
can be controlled by the Kullback-Leibler divergence.\\

One may wonder whether our conclusions, obtained for a rather trivial model from the biological point
of view has any relevance for the analysis of real neurons statistics. We would like to point that
 the main ingredients making this model-statistics so complex are causality induced by dynamics,
and integration over past events via the leak term. We don't see any reason why the situation should
be simpler for more realistic models. Especially, what makes the analysis tractable is the reset
of the membrane potential to a constant value after firing, inherent to IF models, and rather
unrealistic in real neurons. Thus, the memory effects could be even worse in realistic neurons,
with a difficulty to extract, from a thorough mathematical analysis, the relevant times scales for a memory cut-off, as the log of the leak is in the present model. Neverteless, this work is a beginning with a clear need
to be extended toward more realistic models.\\

A natural extension concerns the so-called Generalized Integrate and Fire
models \cite{rudolph-destexhe:06} , which are closer to
 biology \cite{jolivet-rauch-etal:06}.
The occurrence of a post-synaptic potential on synapse $j$, at time $\tjn$,
results in a change of membrane potential. In conductance based models 
this change is integrated in the adaptation of conductances. It has been
shown in  \cite{cessac-vieville:08}, under natural assumptions on
spike-time precision, that the continuous-time evolution of these equations
reduces to the discrete time dynamics. In this case the computation
of the potential corresponding to (\ref{psi}) is clearly more complex but still
manageable.  This case is under current investigations.

One weakness of the present work is that it only considers
stationary dynamics, where e.g. the external current $I_i$
is independent of time. Besides, we consider here an asymptotic
dynamics. However, real neural systems are submitted
to non static stimuli, and transients play a crucial role.
To extend the present analysis to these case one needs the
proper mathematical framework. The non stationarity requires
to handle time dependent Gibbs measures. In the realm of
ergodic theory applied to non equilibrium statistical
physics, Ruelle has introduced the notion of time-dependent
SRB measure \cite{ruelle:99}. A similar approach could be used
here, at least formally.

In neural networks, synaptic weights are not fixed, as in (\ref{DNN}), 
but they evolve with the activity of the pre- and post-synaptic
neuron (synaptic plasticity). This means that synaptic weights evolve
according to spike train statistics, while spike train statistics is
constrained by synaptic weights. This interwoven evolution has been considered in \cite{cessac-rostro-etal:09} under the assumption that spike-train statistics is characterized by a Gibbs distribution. It is shown 
that synaptic mechanism occurring on a time scale which is slow compared to neural dynamics are associated with a variational principle. There is a function, closely related to the topological pressure, which decreases when the synaptic adaptation process takes place. Moreover, the synaptic adaptation has the effect of reinforcing specific terms in the potential, directly related to the form of the synaptic plasticity mechanism.
The interest of this result is that it provides an a priori guess
of the relevant terms in the potential expansion. A contrario, it allows to constrain the spike train statistics of a LIF model, using synaptic plasticity with an appropriate rule which can be determined
from the form of the expected potential.\\

\nid{\small {\bf Acknowledgment:}.} We are indebted to T. Vi\'eville and H. Rostro
for valuable discussions and comments.

\bibliographystyle{unsrt0} 
\bibliography{biblio,odyssee}

\end{document}